# Radio and γ-ray Evidence for the Supernova Origin of High Velocity Cloud Complex M

J.T. Schmelz
0000-0001-7908-6940
USRA, 425 3rd Street SW, Suite 950, Washington, D.C.
jschmelz@usra.edu

G.L. Verschuur
0000-0002-6160-1040
verschuur@aol.com

### Abstract

Using λ21-cm galactic neutral atomic hydrogen data from the HI4PI survey and 0.75-30 MeV γ-ray emission from the Imaging Compton Telescope, we have searched for the origin event that accelerated high velocity cloud Complex M. Radio plots of *l-b*, *l-v*, and *b-v* show a cavity centered at $(l,b) \sim (150°, 50°)$ and extending about ±33°. The best view of the cavity is at a velocity of -25 km/s, which shows a circular cross section on the back (receding) face. Complex M, at -85 km/s, is on the front (approaching) face. The γ-ray emission reveals several minima, the largest centered at $(l,b) \sim (150°, 50°)$ and coincident with the position and extent of the cavity seen in the radio data. Using the know distance to Complex M and assuming that the cavity is spherical, we can bootstrap the distance to the original, explosive source of the cavity, D = 307 pc, calculate the radius of the cavity, R = 166 pc, and approximate the expansion velocity, $V_E \sim 40$ km\s, of the cavity. The total energy of the expanding cavity is $3.0 \pm 1.0 \times 10^{50}$ ergs, well within the range of a single supernova. These results indicate that this explosion took place about four million years ago. As the blast wave from this supernova propagated outwards, it began to sweep up interstellar gas and carved out the Local Chimney, a low-density extension of the Local Bubble that reaches into the galactic halo.

**Introduction**

Early models for the origin and energy of interstellar high-velocity (HV) clouds of neutral hydrogen were summarized by Oort (1966). Promising candidates that grew out of these first discussions include the possibilities that these clouds might be condensations in the hot galactic halo (Oort 1970), part of the outer spiral structure (Verschuur 1973), return flows in a galactic fountain (Bregman 1980), infall to the Galaxy (Mirabel 1982), and fallout from the Perseus supershell (Verschuur 1993). With the notable exception of the Magellanic Stream, which was torn out of the Magellanic Clouds in a tidal interaction with the Milky Way, the debate over the origin stories for HV clouds continues. Please see Verschuur (1975) for an early review and van Woerden et al. (2004) for subsequent developments.

New research suggests that the HV cloud known as MI may be the result of a supernova that blew up about 100,000 years ago (Schmelz & Verschuur 2022). Although this model class was considered by Oort (1966), it was rejected based on the evidence available at the time from optical absorption studies that placed the HV clouds at kiloparsec distances, thereby making the energy considerations unlikely. Schmelz & Verschuur (2022) found that low-velocity (LV) hydrogen data reveal a cavity surrounding MI. Although the cavity itself is neither unique nor rare, its alignment with MI was unexpected because of the traditional view that LV gas is in the solar neighborhood and HV gas is in the galactic halo. For decades, there had been no good reason to connect the two.

The yellow giant star, 56 Ursae Majoris (56 UMa), is part of a single-line spectroscopic binary system that sits in the middle of the MI cavity identified by Schmelz & Verschuur (2022). Its invisible companion had been tagged as a white dwarf (see, e.g., Jorissen et al. 2019 and references therein), but new results combine radial-velocity data with proper motions from Hipparcos and Gaia to re-identify it as a neutron star (Escorza et al. 2022). Schmelz & Verschuur (2022) proposed that this neutron star might be the remains of the supernova that propelled MI to the observed high velocity. The distance to the star is 163 pc, making the distance to MI about 150 pc, close enough that the mass and energy of MI are easily in line with the expectations of a single supernova. Please note: although traditional absorption-line studies place Complex M at kpc distances (Danly et al. 1993), these results are no longer considered valid due to the pervasive filamentary nature of the HI gas (McClure-Griffiths et al. 2009; Peek et al. 2011; Winkel et al. 2016; Martin et al. 2015).

Verschuur & Schmelz (2022) found that several HV features, including MI, MIIa, and MIIb, are components of a long, arched filament. They used maps at different velocities, results from Gaussian analysis, and observations of high-energy emission to make a compelling case that the MI cloud and the arched filament are physically interacting. They could then bootstrap the MI distance to that of the entire structure, proposing that the distance to Complex M is also of order 150 pc. They estimate a mass of about 120 solar masses and an energy of $8.4 \times 10^{48}$ ergs. Integrating over $4\pi$ steradians, the total energy for a spherically symmetrical explosion is approximately $1.9 \times 10^{50}$ ergs, well within the energy budget of a typical supernova.

Schmelz & Verschuur (2022) determined the distance to MI, 150 pc, based on its relationship with the 56 UMa binary system. Verschuur & Schmelz (2022) used this result and the physical connection between MI and other nearby anomalous-velocity features to bootstrap the distance to Complex M, also 150 pc. In this paper, we take the next step in this analysis - looking for evidence of a supernova origin of Complex M.

**Analysis**

The radio data used here are from the λ21-cm galactic neutral atomic hydrogen HI4PI survey (Ben Bekhti et al. 2016), which combines the northern hemisphere data from the Effelsberg-Bonn Survey (EBHIS; Winkel et al. 2016) and the southern hemisphere data from the third revision of the Galactic All-Sky Survey (GASS; McClure-Griffiths et al. 2009). HI4PI has an angular resolution of $\Theta_{FWHM}$ = 16.2 arcmin, a sensitivity of σ = 43 mK, and full spatial sampling, 5 arcmin in both galactic *l* and *b*.

The overall morphology of the area described in this paper is shown in the *l-b* maps of Fig. 1. We created a library of images of the HI brightness from the original HI4PI data cube at 5 km/s intervals in the range from -185 to +20 km/s with a bandwidths of 5 km/s. Examples at several velocities were chosen to highlight some of the crucial morphological details relevant to our analysis.

Fig. 1a shows an *l-b* map at a velocity of -25 km/s. Although there is a lot of complex structure in this map, the outlines of a cavity centered at $(l, b) \sim (150°, 50.°)$ are clearly visible. (Note: The aspect ratio of the rectangular coordinates has been adjusted to highlight the outlines of the cavity.) This position is close to a well-known region of exceptionally low HI column densities (Lockman et al. 1986), but this cavity is much larger. Although the edges of the structure are blurred by foreground and background features, the extent is about 90° to 210° in longitude and 25° to 75° in latitude.

This structure is not as distinct in Fig. 1b at -50 km/s, but the greater extent of the cavity, especially at higher latitudes, is apparent. The cavity appears to reach its maximum extent at this velocity, with its high latitude boundary marked by a narrow filament that extends from $(l, b) = (85°, 50°)$ to $(l, b) = (60°, 240°)$ at the edge of the map shown here. There are major foreground and background structures in this image that prevent us from getting a clean view of the cavity. Perhaps the most dominant is the well-known intermediate-velocity (IV) Arch, which crosses the image from about at $(l, b) = (100°, 50.°)$ to $(l, b) = (200°, 50.°)$.

Fig. 1c at -85 km/s shows the Complex M arched filament described by Verschuur & Schmelz (2022), which extends from about $(l,b) = (105°, 50°)$ to $(l,b) = (200°, 45°)$. The brighter features within its length, which appeared to be isolated "clouds" in maps from old surveys, are now seen as local enhancements in a long, twisted filament. This revised view is the result of the improved sensitivity, resolution, and dynamic range available with HI4PI and other recent HI surveys. Parts of the IV Arch are also visible here extending northward at $l = 120°$ from $b = 25$ to $60°$.

Different perspectives on the structure and extent of the cavity are revealed in the position-velocity maps from the HI4PI data cube. For example, Fig. 2a shows an *l-v* plot at $b = +50°$ produced by averaging data over 0.°1 in latitude. The contours were chosen to highlight the cavity centered on a velocity of about -25 km/s (see Fig. 1a) and extending from about 100° to 190° in longitude. The map reveals a lot of complicated structure, including a ridge of emission around -60 km/s that forms part of the IV Arch. Fig. 2b shows a *l-v* plot at $b = +81°$ which cuts across the top of the cavity at $l = 150°$. The filament-like ridge in Fig. 1b is evident here from $l = 110°$ to $l = 170°$. This image shows that the emission at $l > 180°$ belongs to a different feature in the third quadrant of galactic longitude. Fig. 3 shows a *b-v* plot from the HI4PI data cube at $l = 150°$ produced by averaging data over 0.°1 in longitude. As in Fig. 2a, the contours were chosen to highlight the cavity centered on a velocity of about -25 km/s and extending from about 30° to 70° in latitude.

If the cavity and the surrounding disrupted gas were the result of an old supernova, remaining neutron star might be the "smoking gun," but without a bright binary companion, a pulsar signature, or evidence of mass transfer, it would be all but invisible to us. A map of the total HI column density integrated over the velocity range -100 to +25 km/s revealed several local minima, the deepest at $(l, b) = (152.°25, 52.°0)$ with column density of $5.0 \times 10^{18}$ cm$^{-2}$. A search of the SIMBAD data base (https://simbad.u-strasbg.fr/simbad/sim-fcoo) near these local minima revealed no likely candidates for the original source of the explosive energy. There is, however, high-energy emission from this area. An extended, nebulous cloud of X-rays was detected as part of the ROSAT 1/4keV survey (Herbstmeier et al. 1995). Diffuse γ-ray emission in the range 0.75-3 MeV was also observed in this region with COMPTEL (Blom et al. 1997).

Fig. 4a shows the γ-ray emission in the range 1-30 MeV for entire Northern Galactic hemisphere from the COMPTEL survey described by Schonfelder et al. (1993). There are several extended minima in the γ-ray emission, the largest centered at around $(l, b) = (150°, 50°)$ and coincident with the position and extent of the cavity seen in the HI4PI radio data in Fig. 1a. There are no comparable structures in the Southern Galactic hemisphere.

Fig. 4b shows a detailed view of this minimum from $l = 60°$ to $240°$ and $b = 20°$ and $83°$. The Complex M cavity centered at $(l, b) \sim (150°, 50°)$ is clearly visible. Although the edges do not form a smooth circle, the extent is about 110° to 210° in longitude and 20° to 70° in latitude. The HI contours from the cavity shown in Fig. 1a are overplotted. There are two γ-ray peaks in the cavity at about $(l, b) = (163°, 49°)$ and $(l, b) = (138°, 48°)$, which may or may not be foreground/background sources.

These high-energy emissions are often attributed to non-thermal electron bremsstrahlung arising from HV cloud interactions with matter at the halo-disk interface, but since this border is absent in the Local Chimney (see Discussion), a different explanation is required. Rather, they may be a manifestation of the explosive process that evacuated the cavity and generated the HV gas. As the supernova blast wave pushes into the rarefied interstellar medium, the shock waves at the boundary can produce γ-rays and cosmic rays. These high-energy emissions could point the way to understanding the energetics of the phenomenon that created Complex M.

**Discussion**

Fig. 5 shows the suggested geometry of the expanding cavity seen in the HI4PI λ21-cm galactic neutral atomic hydrogen data discussed above. We make the simplifying assumptions that the cavity is spherical in this coordinate system and is expanding with a constant velocity. The first triangle in Fig. 5a, triangle OTS, connects the Sun at the origin (O), the tangent point (T) that defines a right angle, and the source (S) of the initial explosion. The geometry requires information from Fig. 1b, which shows the greatest extent of the cavity where the cross section in the plane of the sky is a great circle. The angle, angle TOS, is the radius of this great circle, the latitude of the maximum extent minus the latitude of the center, or angle TOS = 83° - 50° = 33°. Since triangle OTS is a right triangle, $\sin 33° = ST/OS = R/D = 0.54$, where R is the radius of the cavity and D is the distance to the source of the explosion.

The second triangle in Fig. 5a, triangle OMS, connects MI to the Sun at the origin (O) and the source (S) of the explosion. triangle OMS is not a right triangle but shares a baseline with triangle OTS. The geometry requires information from Fig. 1c, which shows the arc of Complex M which we use to define a circular cross section (not a great circle) on the front surface of the expanding shell. The angle, angle MOS, is the radius of this circle, the latitude of MI minus the latitude of the center, or angle MOS = 65° - 50° = 15°. Using the Law of Sines, angle OMS = 151°.5, R = 166 pc, and D = 307 pc.

Next, we derive an approximation of the expansion velocity of the shell using the red triangle in Fig. 5b. The systemic velocity of a spherical shell expanding at a uniform velocity is determined by its maximum extent, -50 km/s (Fig. 1b). Complex M is on the front face of the shell (but not at the apex) approaching us at 85 km/s or at $V_{Ey} = V_M = 35$ km/s plus the systemic velocity. The radius, $R_x = R_M$, of the Complex M circle (not a great circle), is $\sin 15° = R_M/D$.

$R_M = D \times \sin 15° = 307$ pc $\times 0.26 = 80$ pc.

For the right triangle in Fig. 5b, $\cos \theta_M = R_M/R = 80/166 = 0.48$ and $\theta_M = 61°.3$. To find the expansion velocity, $V_E$, of the cavity, we use the same triangle, except for velocity. $\sin \theta_M = V_M/V_E$, and solving for $V_E$ gives:

$V_E = V_M/\sin 61.°3 = 35$ km/s $/ 0.877 \sim 40\backslash$ km/s.

Doing the same calculation for the Cavity circle on the back face of the shell (Fig. 1a), which is approaching us at 25 km/s or at $V_{Ey} = V_C = 25$ km/s plus the systemic velocity. The radius, $R_x = R_C$, is $\sin 25° = R_C/D = 129$ pc. For the right triangle in Fig. 5b, $\cos \theta_C = R_C/R = 129/166 = 0.777$ and $\theta_C = 39°$. To find the expansion velocity, $V_E$, of the cavity, we use the same triangle, except for velocity. $\sin \theta_C = V_C/V_E$, and solving for $V_E$ gives:

$V_E = V_C/\sin 39° = 25$ km/s $/ 0.63 \sim 40\backslash$ km/s.

The fact that the front and back faces of the shell are expanding at the same velocity indicates that the spherical symmetry of the cavity is a reasonable approximation. If the derived expansion velocity of 40 km/s were constant, then the age of the shell would be about four million years. This should be considered an upper limit since the velocity decreases with time. Therefore, the event that evacuated the cavity seen in Fig. 1a, accelerated Complex M to the observed high velocities, and produced the γ-rays seen in Fig. 4b occurred less than or about equal to 4 million years ago. The approximate position on the sky of the possible neutron star that resulted from the explosion (source S) is simply the center of the cavity seen in Fig. 1a, $(l, b) = (150°, 50°)$. The distance to this neutron star is calculated from the trigonometry of the triangles shown in Fig. 5a, D ~ 307 pc. The radial velocity of this neutron star coincides with the greatest extent of the cavity, about -50 km/s.

Without the original radial velocity of the supernova progenitor, Verschuur & Schmelz (2022) could only estimate the energy required to accelerate the matter of Complex M from zero velocity. Here the HI4PI data allow us to estimate the average HI column density for the filament segments at -25, -55 and -85 km/s (Fig. 1). Making the standard assumption that the depth is about equal to the width of the filaments, the average derived density is $1.0 \pm 0.3$ cm$^{-3}$. Using the expansion velocity calculated above, the average energy of the shell is of order $3.0 \pm 1.0 \times 10^{50}$ ergs, well within the energy budget of a typical supernova ($10^{51}$ ergs).

The location of this Complex M cavity places it in the Local Chimney (Welsh et al. 1999), a low-density extension of the Local Bubble (Zucker et al. 2022) that reaches all the way into the galactic halo. The bubble and chimney were mapped in 3D using interstellar absorption characteristics of the NaI D-line doublet at 5890 Å by Lallement et al. (2003) and references therein. They targeted over 1000 stars within 350 pc of the Sun whose distances were determined by Hipparcos parallax measurements. Observations indicate that the Local Bubble appears to extend to 100-200 pc in all directions and is surrounded by an irregular, higher-density gas boundary near the galactic plane.

The lightest grey shading in Fig. 6 shows a cross section of the Local Chimney at longitude 150° adapted from Lallement et al. (2003). The distance to the source of the explosion (star), D = 307 pc, and the extent of the Complex M cavity (red arc), R = 166 pc, are also shown to scale. The darker shading along the 30° direction at the position of the red arc may be the HI in the lower right quadrant of Fig. 1a. This denser gas may hinder the uniform expansion of the shell in this direction. In fact, any density gradient in the interstellar medium may lead to a non-spherical expansion of the cavity.

The size, location, and age of this cavity indicate that it might be responsible, at least in part, for the formation of the Local Chimney. The scenario is as follows - a star at position S in Fig. 5 went supernova about four million years ago. As the blast wave propagated outwards and moved through interstellar space, it began to sweep up gas via the snowplow effect (Spitzer 1978). The mass on or near the surface of the shell, which includes Complex M, is a combination of the mass from the original star system and the mass from the surrounding interstellar matter.

If the blast wave had accelerated atomic gas, one might expect to see more than a semi-complete rim (Fig. 1a), an arc of maximum extent (Fig. 1b), and a fast-moving filament (Fig. 1c). Where is the expanding HI in the -50 ± 40 km/s range, that is, the near- and far-side surfaces? The best way to address this important point is to compare what we observe with the more ideal version of a supernova explosion/cavity creation, where a spherically symmetrical explosion would push a uniform-density medium into a shell, and nothing would obscure our view of these phenomena. Reality, as expected, is more complex.

The pre-supernova mass loss from the star at position S was almost certainly clumpy (Hamann et al. 2008; Smith 2014). Some of the features in Complex M may very well be a collection of these surviving clumps that rode the blast wave to achieve their observed high velocities. The explosion is also pushing a complex, clumpy, filamentary medium. This is why the rim seen in Fig. 1a is notable. Here, we get a close approximation to the ideal case, where the combination of the snowplow effect and limb brightening combine to make this rim stand out. At other velocities, however, the hydrogen shows the clumpy, non-uniform structure that seems to prevail, at least inside the Local Chimney. As seen in Figs. 1-3, the IV gas along this line of sight is complex, and the IV Arch dominates many of the *l-b, l-v,* and *b-v* plots. Although the distance to the IV Arch is unknown, traditional models would place it between the LV and HV gas, that is, in the same region of space as our expanding shell. It may be behind, in front of, or even part of the supernova shell (but this is beyond the scope of this paper).

This brings up a challenge related to the modern study of interstellar HI gas. What appeared to be isolated "clouds" in maps from old surveys might now be seen as parts of filaments as sensitivity, resolution, and dynamic range improve. Where clouds had clear boundaries and were isolated on the sky, filaments connect features in both position and velocity space. Defining the boundaries is not necessarily easy, even with psudo-3D, *l-b-v* data. Although the IV Arch dominates our images, it does not appear, at least so far, to add to or subtract from the supernova scenario.

## Conclusions

Using λ21-cm galactic neutral atomic hydrogen data from HI4PI (Bekhti et al. 2016) and 1-30 MeV γ-ray emission from COMPTEL (Schonfelder et al. 1993), we have searched for the origin event that accelerated Complex M to its high velocities.

Radio plots of *l-b, l-v*, and *b-v* show a cavity centered at (*l, b*) ~ (150°, 50°) with a radius of about 33°. The best view of the cavity is at a velocity of -25 km/s, which shows a circular cross section on the back (receding) face of this cavity. Complex M is on the front (approaching) face of this cavity.

The Complex M cavity is also seen in the γ-ray data from COMPTEL (Fig. 4). As the supernova blast wave pushes into the rarefied interstellar medium, the shock waves at the boundary can produce γ-rays and cosmic rays.

Using the know distance to Complex M and assuming that the cavity is spherical, we can bootstrap the distance to the original source of the cavity, D = 307 pc, and calculate the radius of the cavity, R = 166 pc.

Assuming that Complex M is on the front face of the expanding, spherical shell of gas, we can calculate the expansion velocity, $V_E$ ~ 40 km/s. Repeating the calculation for the rim of gas seen in Fig. 1a and assuming it is on the back face of the shell gives us the same expansion velocity. This indicates that our assumption of a spherical shell is reasonable.

The expansion velocity and extent of the shell indicate that the original explosive event, a supernova, took place about four million years ago. The total energy required to move the mass at a velocity of 40 km/s is $3.0 \pm 1.0 \times 10^{50}$ ergs, within the energy budget of a typical supernova ($10^{51}$ ergs).

As the blast wave from this supernova propagated outwards, it began to sweep up interstellar gas and carve out the Local Chimney, a low-density extension of the Local Bubble that reaches all the way into the galactic halo.

## Acknowledgments

We acknowledge the use of NASA's SkyView facility (http://skyview.gsfc.nasa.gov) located at NASA Goddard Space Flight Center. We are grateful to T. Dame for providing us with his MacFits software that seamlessly allows us to unravel data cubes, to J. Kerp for providing the *HI4PI* data, and to W.B. Burton for helpful advice.

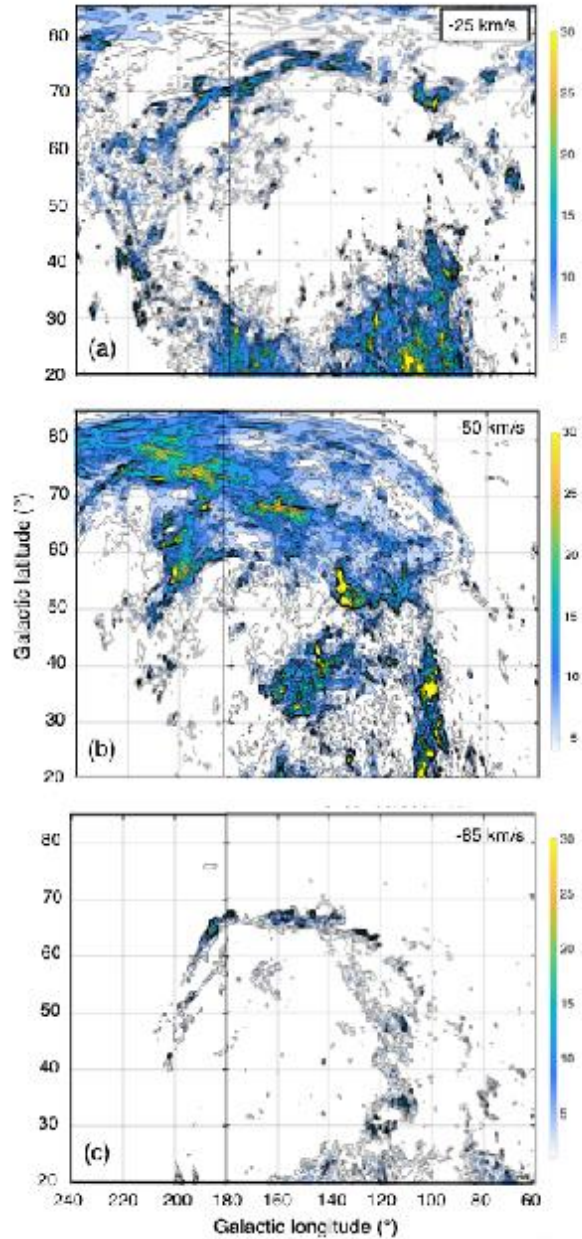

Fig. 1 Area maps of the HI emission from *HI4PI* data at the velocities indicated with a bandwidth of 5 km/s. (a) The *l-b* map at -25 km/s showing the outlines of a cavity centered approximately at (*l, b*) = (150°, 50°) and extending from about 90° to 210° in longitude and 25° to 75° in latitude. (b) The *l-b* map at -50 km/s showing the greater extent of the cavity, especially at higher latitudes, reaching up to about 83°. Emission from the IV Arch extends from about (*l,b*) = (100°, 50°) to (*l, b*) = (200°, 80°). (c) The *l-b* map at -85 km/s shows the Complex M arched filament, which stretches from about (*l, b*) = (90°, 40°) to (*l, b*) = (200°, 40°). Parts of the IV Arch are visible around *l* =120°. Legends are in Kelvins.

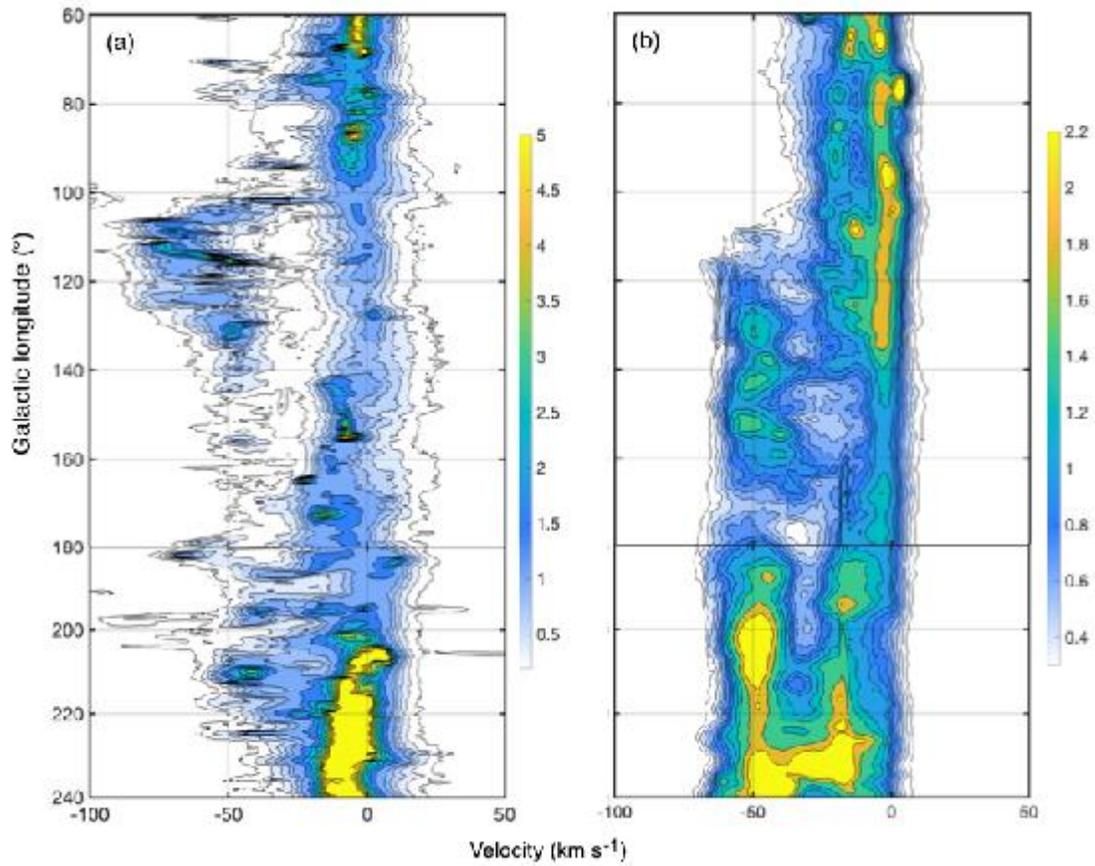

Fig. 2 a) The *l-v* plot from the HI4PI data cube at $b = +50°$ produced by averaging over 1° in latitude. The contours were chosen to highlight the cavity centered on a velocity of about -25 km/s, which extends from about 90° to 210° in longitude. It shows a lot of complex structure, including a ridge of emission around -60 km/s that forms part of the IV Arch. b) The *l-v* plot at $b = +81°$ marks the top of the cavity at $l = 150°$ and stretches from about 120° to 170° in longitude. The ridge of emission at $l > 180°$ belongs to a distinctly different feature. This emission at -50 km/s centered at $l = 150°$ determines the greatest extent of the cavity seen in Fig. 1b. Legends in Kelvins.

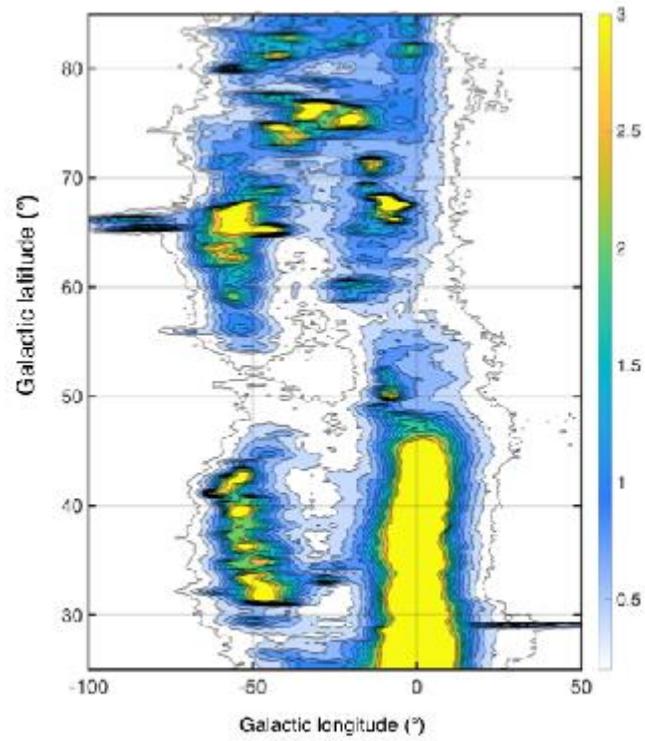

Fig. 3 The *b-v* plot from the HI4PI data cube at $l = 150°$ averaged over 1° in longitude. The contours were chosen to highlight the cavity centered on a velocity of about -50 km/s that extends from about 30° to 70° in latitude. IV Arch emission is present at -60 km/s.

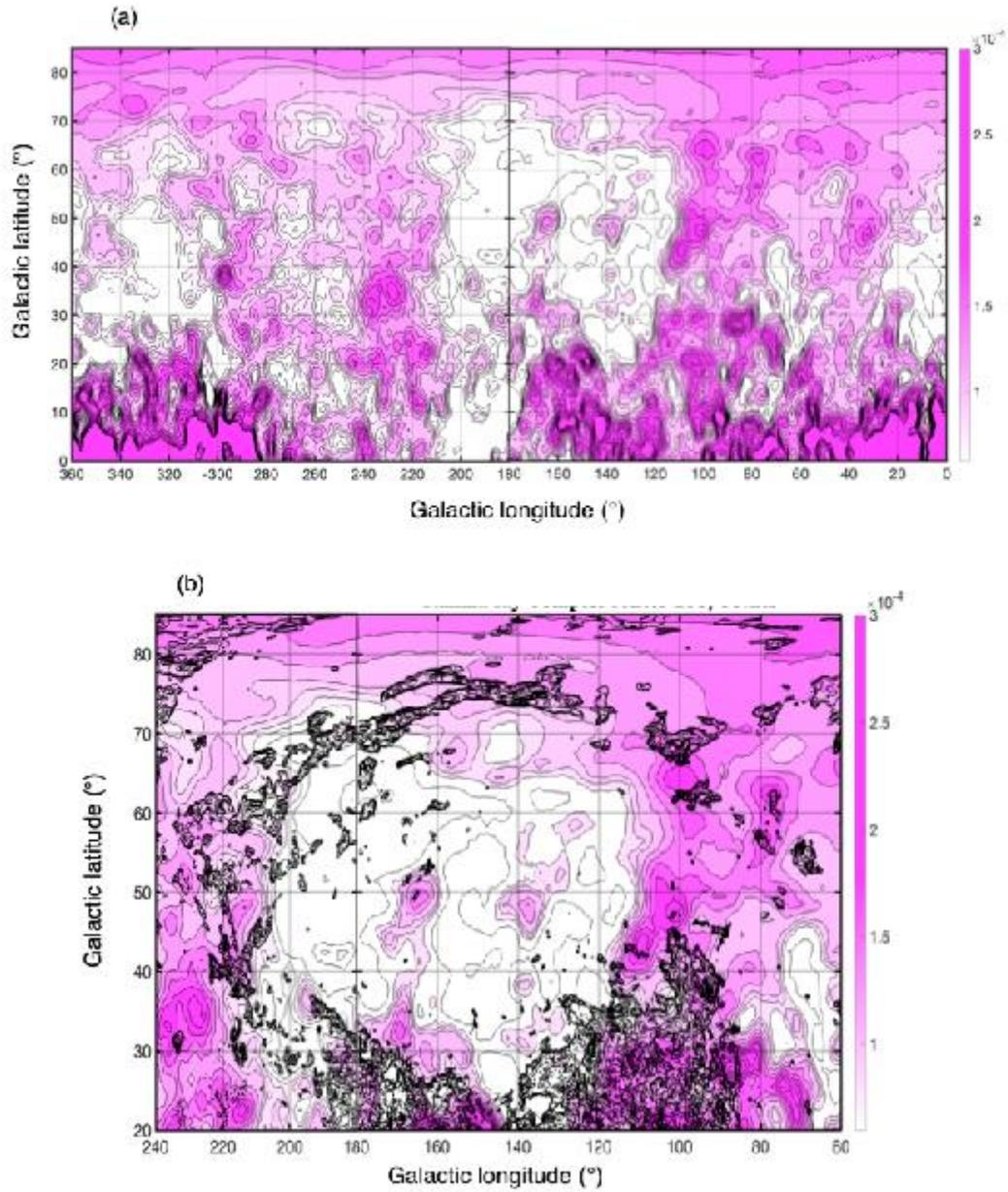

Fig. 4 (a) Diffuse γ-ray emission from the COMPTEL survey over the full northern Galactic hemisphere. The large minimum in the center of the plot marks the Complex M cavity. (b) Close up of the Complex M cavity with the COMPTEL γ-rays (magenta) and the λ21-cm HI emission (contours) at -25 km/s from Fig 1a. The legends are in units of cts/s/steradian.

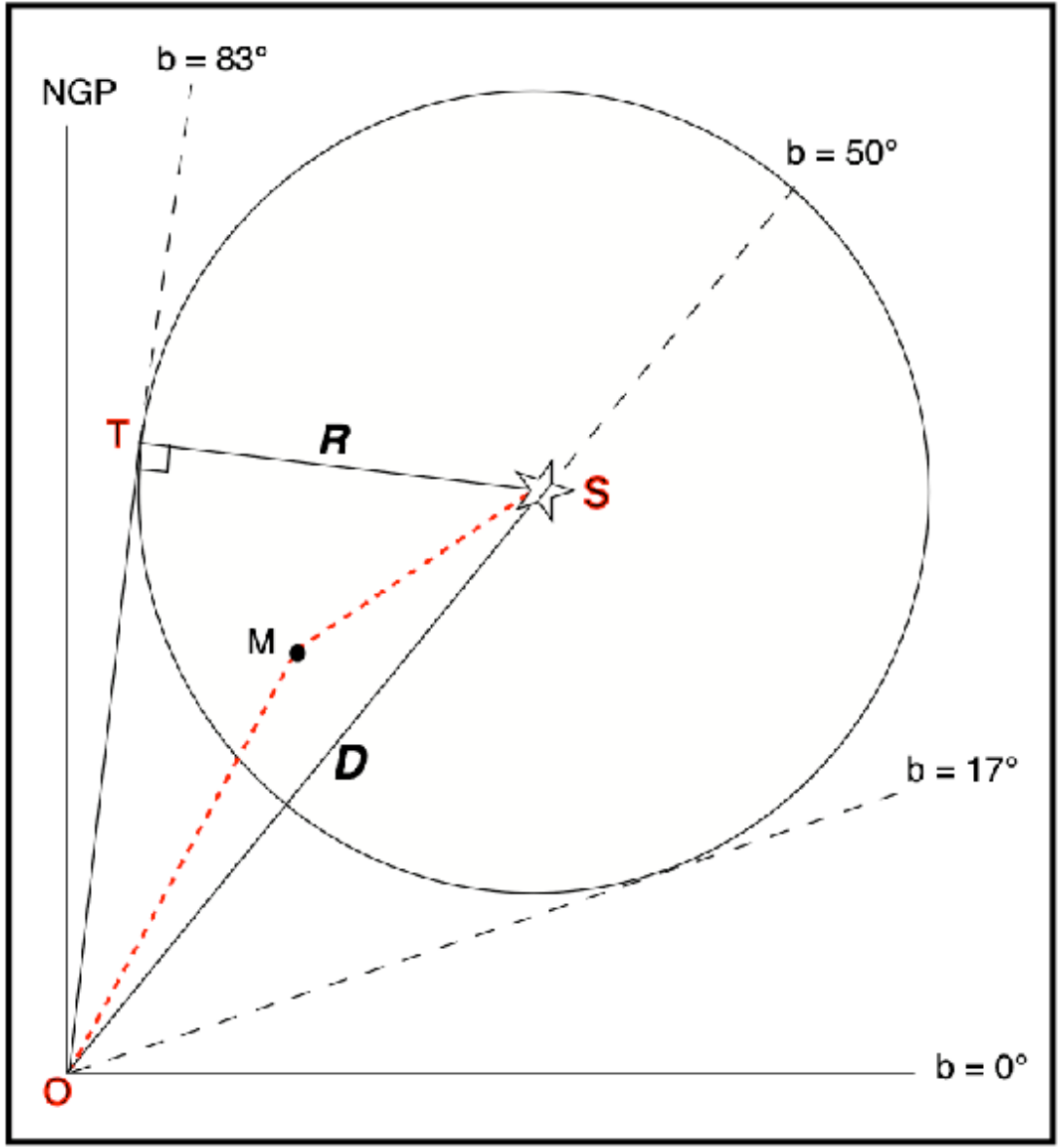

Fig. 5a A schematic of the geometry of the region seen in Fig. 1 where we have made the simplifying assumption that the cavity has a spherical cross-section at $l = 150°$. The triangle, triangle OTS, connects the Sun at the origin (O), the tangent point (T) at the maximum extent of the cavity, and the source (S) of the initial explosion. The tangent point defines a right angle and angle TOS $= 33°$ (see text). The second triangle, triangle OMS, connects MI to the Sun at the origin (O) and the source (S) of the explosion. Triangle OMS is not a right triangle, but it shares a baseline with triangle OTS. Angle MOS $= 15°$.

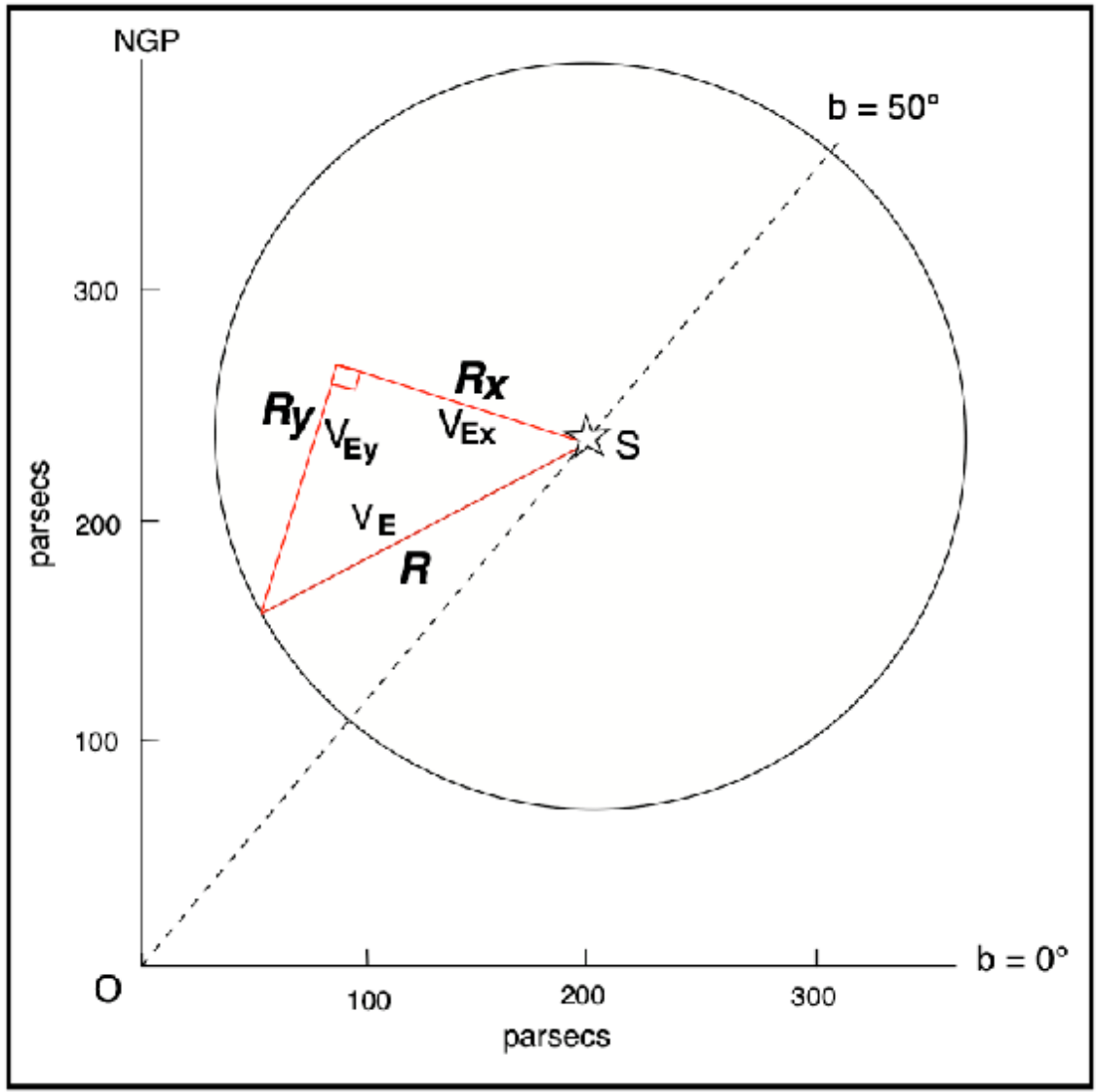

Fig. 5b A schematic showing the derived geometry of the shell to scale at $l = 150°$. The red triangle has sides R-$R_x$-$R_y$, where R is the radius of the cavity. This triangle also has sides $V_E$-$V_{Ex}$-$V_{Ey}$, where $V_E$ is the expansion velocity of the cavity.

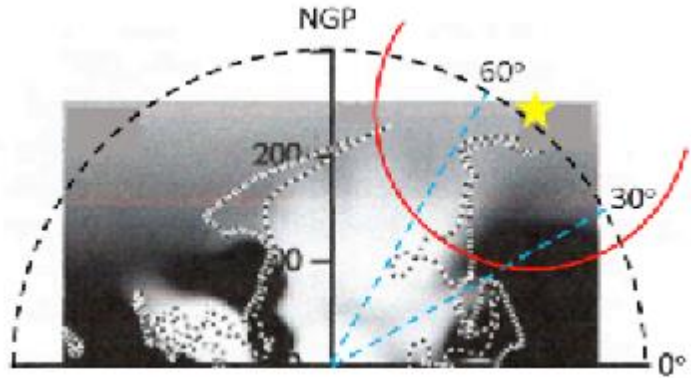

Fig. 6 The outlines of the Local Chimney adapted from Fig. 8 of Lallement et al. (2003). This is the cross-section at $l = 150°$ with distances along the two axes shown in pc. The distance to the source of the explosion (star), D = 307 pc, and the extent of the Complex M cavity (red arc), R = 166 pc, are also shown to scale.